\documentclass[]{./aa}
\usepackage{amsmath, amssymb, upgreek, ulem}

\usepackage{graphicx}
\usepackage{txfonts}
\usepackage{natbib}
\begin{document}
   \title{No evidence for planetary influence on solar activity}

   %\subtitle{}

   \author{R.H.~Cameron\inst{1}
          \and M.~Sch\"ussler\inst{1}}

   \institute{Max-Planck-Institut f\"ur Sonnensystemforschung,
              Max-Planck-Str. 2, 37191 Katlenburg-Lindau, Germany\\
              \email{cameron@mps.mpg.de,schuessler@mps.mpg.de}
                       }
   \date{\today}

\abstract{Recently, Abreu et al. (2012, A\&A. 548, A88) proposed a
  long-term modulation of solar activity through tidal effects exerted
  by the planets. This claim is based upon a comparison of
  (pseudo-)\-periodicities derived from records of cosmogenic isotopes
  with those arising from planetary torques on an ellipsoidally deformed
  Sun.}
 {We examined the statistical significance of the reported similarity
  of the periods.}
 {The tests carried out by Abreu et al. were repeated with
  artificial records of solar activity in the form of white or red
  noise.  The tests were corrected for errors in the noise definition as
  well as in the apodisation and filtering of the random series.}
 {The corrected tests provide probabilities for chance coincidence that
  are higher than those claimed by Abreu et al. by about 3 and 8
  orders of magnitude for white and red noise, respectively.
  For an unbiased choice of the width of the frequency
  bins used for the test (a constant multiple of the frequency
  resolution) the probabilities increase by another two orders of
  magnitude to 7.5\% for red noise and 22\% for white noise.}
 {The apparent agreement between the periodicities in records of
 cosmogenic isotopes as proxies for solar activity and
 planetary torques is statistically insignificant. There
 is no evidence for a planetary influence on solar activity.}
   \authorrunning{Cameron \& Sch\"ussler}
   \titlerunning{Apparent planetary influence on solar activity}
   \maketitle
%
%________________________________________________________________

\section{Introduction}
\label{sec:intro}
There is a long record of attempts to associate periodicities in the
level of solar activity with the orbits of the planets. All of these
eventually failed rigorous statistical tests \citep{Charbonneau:2002},
which is not surprising in view of the extreme tininess of the physical
effects \citep[e.g.,][]{Callebaut:etal:2012}.  

Recently, \citet[][hereafter A2012]{Abreu:etal:2012} made
a new attempt in this direction by comparing periodicities detected in
the records of cosmogenic isotopes $^{10}$Be and $^{14}$C (or quantities
derived from them) as proxies for solar activity in the past 9400 years
with those of the torque exerted on a thin shell of an ellipsoidally
deformed Sun%
\footnote{The origin of such a localised deformation of the solar mass
  distribution is left obscure by the authors. They refer to results from
  helioseismology indicating a prolate tachocline, but the latter is a
  feature of differential rotation, which could at most slightly modify
  the anyway minuscule rotational deformation of the solar
  equipotential surfaces.}.
They found coincidences between selected periodicities in the planetary
torque and the level of cosmogenic isotopes.  After assessing the
statistical significance under the assumption that the level of solar
activity is a realisation of either white or red noise, they interpret
their result as evidence for a planetary influence on long-term
variations of the activity (in their words:``... highly statistically
significant evidence for a causal relationship...''). 

Here we show that the statistical test presented by A2012 to demonstrate
a causal link between the planetary orbits and the level of solar
activity is conceptionally flawed and biased.  Furthermore, their
execution of the test contains severe technical errors.  A corrected test
reveals that the period coincidences reported by A2012 are statistically
insignificant.

\section{The procedure of Abreu et al.}

A2012 chose 5 bands in the period range between 40~yr and 600~yr based
on Fourier transforms of three records related to cosmogenic isotopes
($^{10}$Be, $^{14}$C, and solar modulation potential $\phi$, which are
regarded as proxies of solar activity) and of the calculated time
evolution of the planetary torque.  By construction, each of these 5
bands contains both a spectral peak of the planetary torque series and
of the cosmogenic records.  The details of their procedure involving
7000-yr subsets of the data are described in Appendix A of A2012.

To determine if the similarity between the planetary torques and
cosmogenic records could be due to chance, A2012 generated random $\phi$
records consisting of (what they considered to be) white or red noise
and determined the probability that at least one of the 20 strongest
peaks of the corresponding Fourier spectra falls in each of their
spectral windows. They find very low probabilities, which they interpret
as evidence for a planetary influence on solar activity.

\section{Errors in the statistical test}
There are four distinct errors in the analysis of statistical
significance presented in A2012.

\subsection{Conceptual error}
\label{subsec:logic}

\begin{table*}[ht!]
\caption{Probabilities for one of the top-twenty peaks falling
in spectral windows (cf. Table~2)}
%\begin{center}
\begin{tabular}{c c c c c c c l}
\noalign{\vspace{1mm}}
\hline
\noalign{\vspace{0.5mm}}
Noise & Source & I & II & III & IV & V & all
     \\[0.5ex]
\hline
\rule[2mm]{0mm}{1mm}%

`white' &Abreu et al. (2012)& 0.143 & 0.104 & 0.168 & 0.189 & 0.0011 & 5.04$\times 10^{-7}$
       \\ [0.5ex]
`red'   &Abreu et al. (2012)&  2$\times 10^{-6}$  & 0.001 & 0.454 & 0.547 & 0.091 &$4.61\times 10^{-11}$ 
       \\ [0.5ex]
\noalign{\vspace{1mm}}
\hline
\noalign{\vspace{0.5mm}}
   & Corrections&   &   &   &   &   & 
     \\[0.5ex]
\hline
\rule[3mm]{0mm}{2mm}%
white &noise definition& 0.46 & 0.29 & 0.22 & 0.09 & 0.04 & 1$\times 10^{-4}$
       \\ [0.5ex]
red   &noise definition&  0.02  & 0.06 & 0.31 & 0.34 & 0.24 & 7$\times 10^{-5}$ 
       \\ [0.5ex]
white &apodisation/filtering& 0.59 & 0.40 & 0.32 & 0.13 & 0.06 & 5.2$\times 10^{-4}$
       \\ [0.5ex]
red   &apodisation/filtering&  0.46 & 0.45 & 0.52 & 0.24 & 0.11 & 2.5$\times 10^{-3}$ 
       \\ [0.5ex]
white &window bias& 0.59 & 0.63 & 0.62 & 0.60 & 0.61 & 7.5$\times 10^{-2}$
       \\ [0.5ex]
red   &window bias& 0.46 & 0.68 & 0.74 & 0.92 & 0.94 & 2.2$\times 10^{-1}$ 
       \\ [0.5ex]
%\noalign{\vspace*{2mm}}
\hline
\end{tabular}
\label{tab:noise}
%\end{center}
\end{table*}

In their approach, the authors of A2012 in effect test the following
null hypothesis: ``The occurrence of at least one of the 20 strongest
peaks in the preselected spectral windows is consistent with white or
red noise''. If this null hypothesis could be rejected at a high
significance level, it would only imply that white or red noise can
probably be ruled out as a model for the record of cosmogenic isotopes
(as a proxy for solar activity) in the spectral range considered. This
would not allow to draw any conclusion about a causal relation between
planetary torque and solar activity. Such a relation could only be
supported if virtually {\it all\/} possible statistical models for the
solar activity record that do not have distinct periodicities in the
given spectral range (e.g., other kinds of noise, nonlinear models,
deterministic chaos) could be ruled out in the same way. %
\ifnum 2<1
\footnote{In fact, the shape of the spectrum of the modulation potential
$\phi$ in Figure~5 of A2012 shows that it is probably not realisation of
either white noise or red noise. In particular it has too little power
at periods less than 80 years to (probably) be white noise, and the
spectrum from about 100 years to 600 years is too flat to (probably) be
red noise.}.
\fi
By way of contrast, the failure of the test (inability to reject the
null hypothesis) in only one of these cases is sufficient to show that
the spectral coincidences do not support the presumed causal
relationship. Therefore, the test of A2012 only leads to a clear
conclusion if it fails.

In the subsequent subsections we repeat the test of A2012 and
consider whether the spectrum with periods between 40 and 600 years
generated from white or red noise is consistent with the
observations. Both kinds of noise might not be particularly sensible
physical models for solar activity (or, for that matter, for planetary
effects since the test is symmetric), even over the restricted range of
periods considered, but the aim here is to follow the analysis in A2012
and see if either of these models can be rejected.  Our results show
that the test as defined by A2012 indeed fails, i.e., that the period
coincidences in the solar and planetary records are statistically
consistent with both white and red noise. The contrasting result of
A2012 follows from three errors in their definition and execution of the
test.

\subsection{Error in creating realisations of white and red noise }

\begin{figure*}
\begin{center}
\includegraphics[scale=0.45]{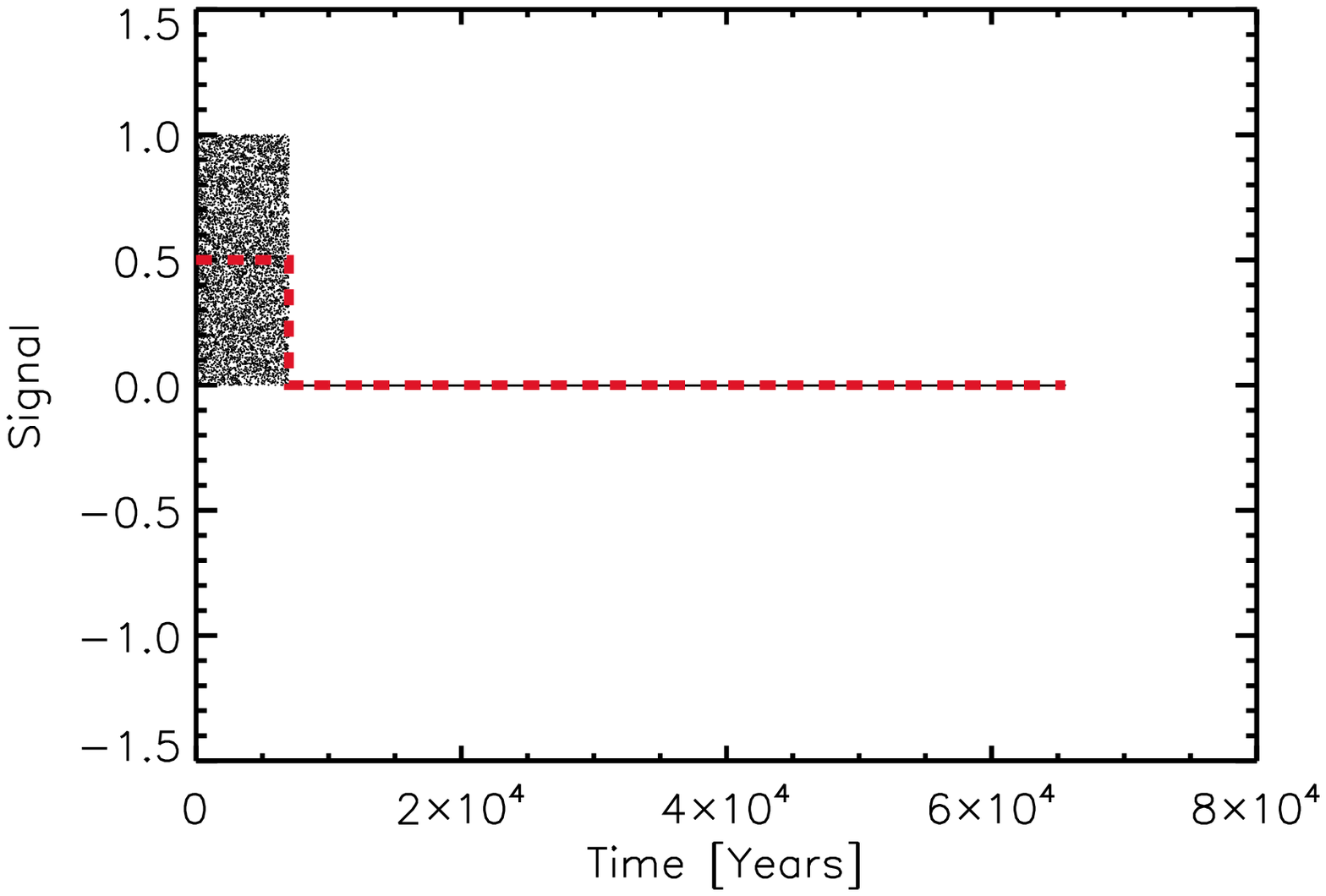}
\includegraphics[scale=0.45]{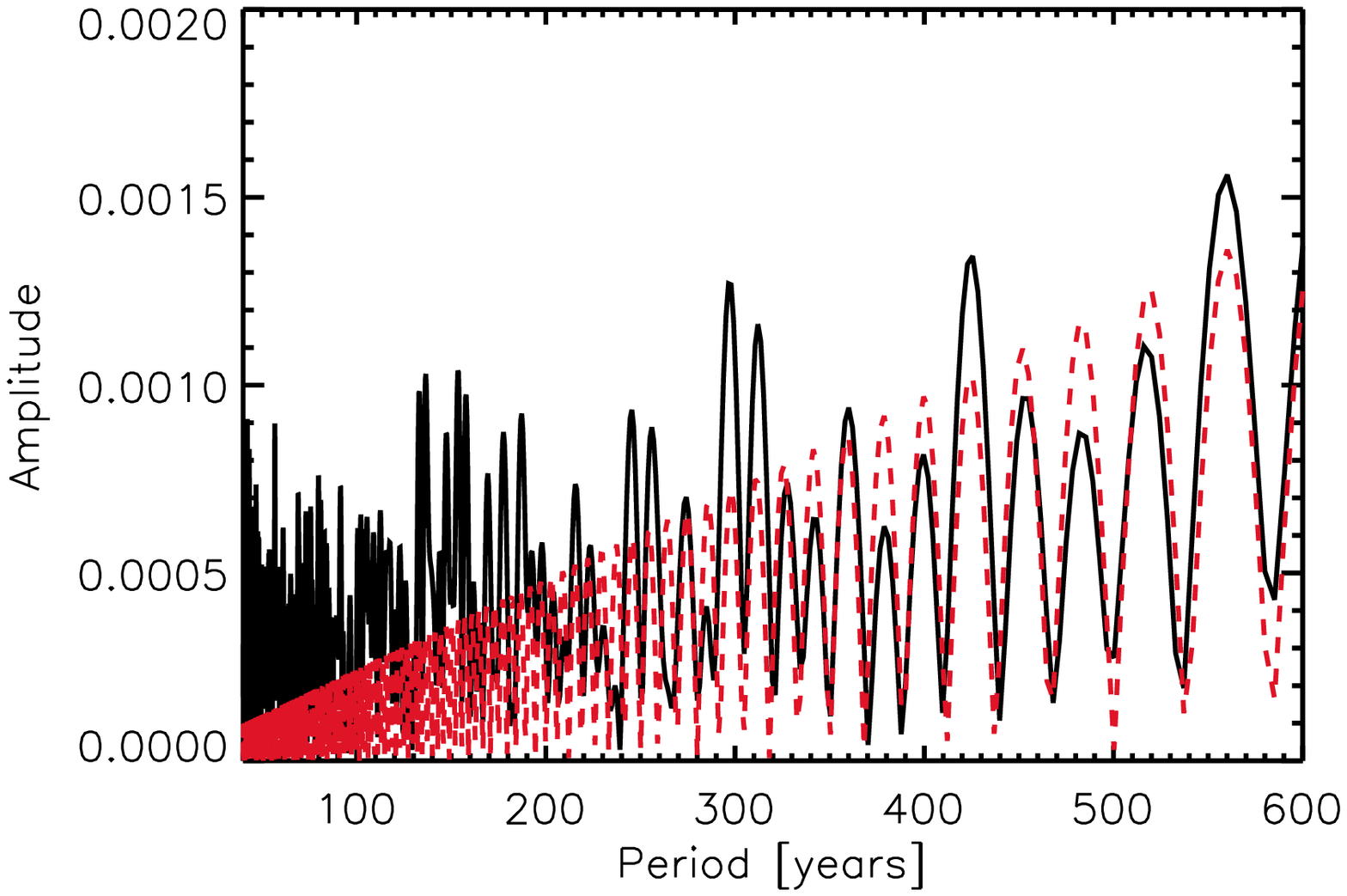}\\
\includegraphics[scale=0.45]{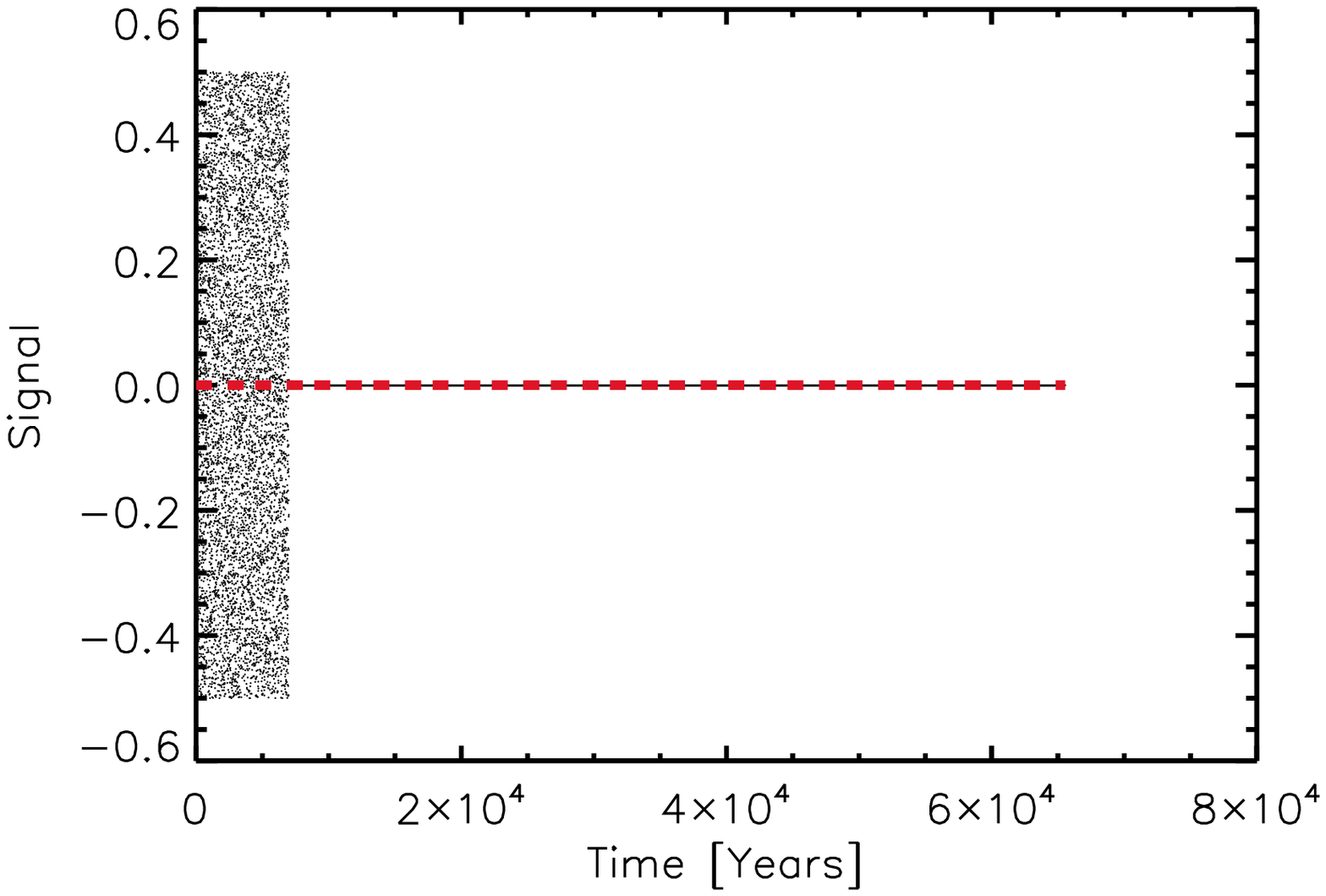}
\includegraphics[scale=0.45]{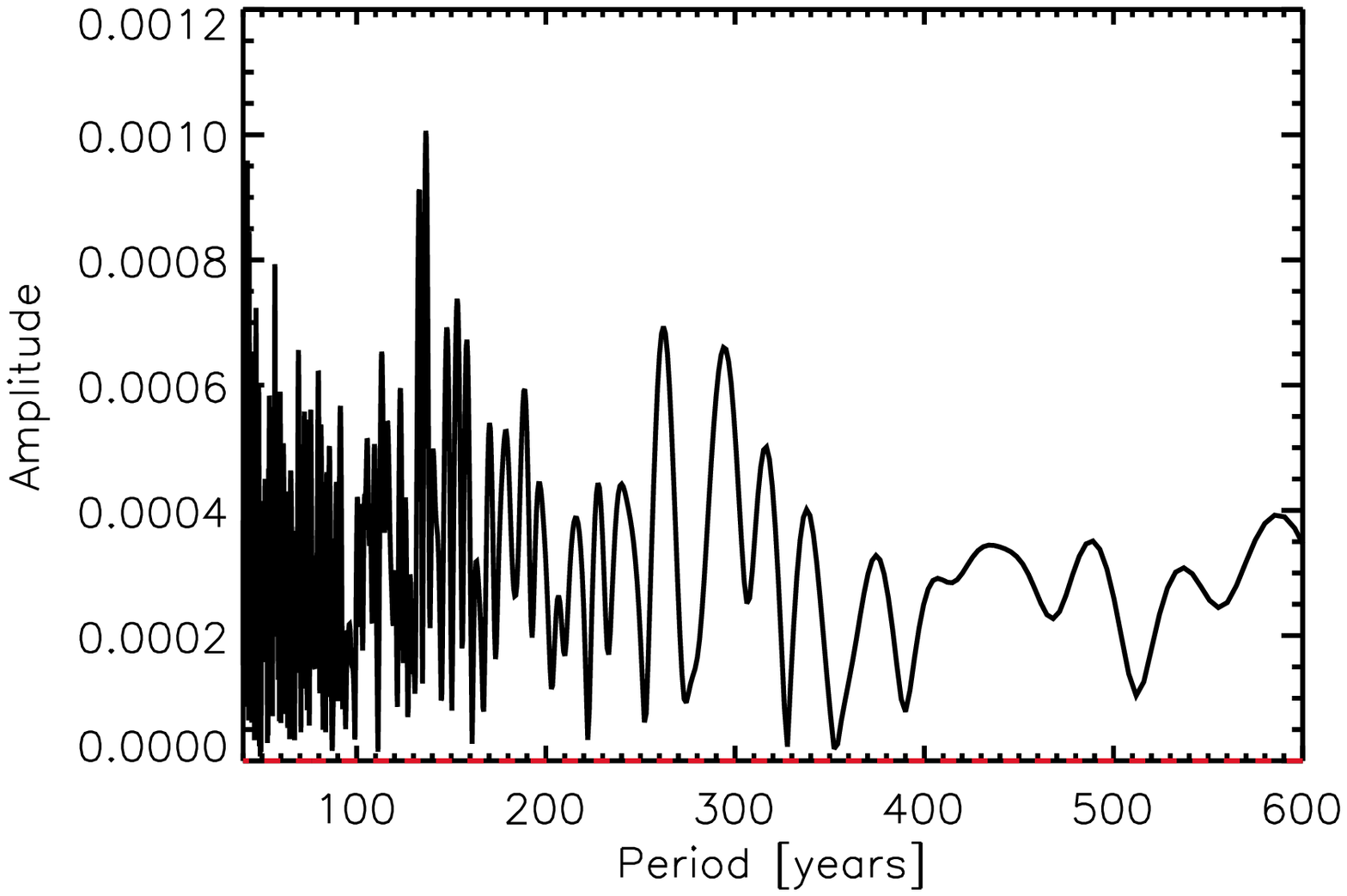}
\caption{`White noise' with zero padding as defined by A2012
         (upper row) and properly constructed (lower row).
         The left panels show the time series while the right
	 panels give the corresponding Fourier spectra. Red dashed
	 lines indicate the systematic component of the signal and
	 its spectrum.}
\label{fig:white}
\end{center}
\end{figure*}

\begin{figure*}
\begin{center}
\includegraphics[scale=0.45]{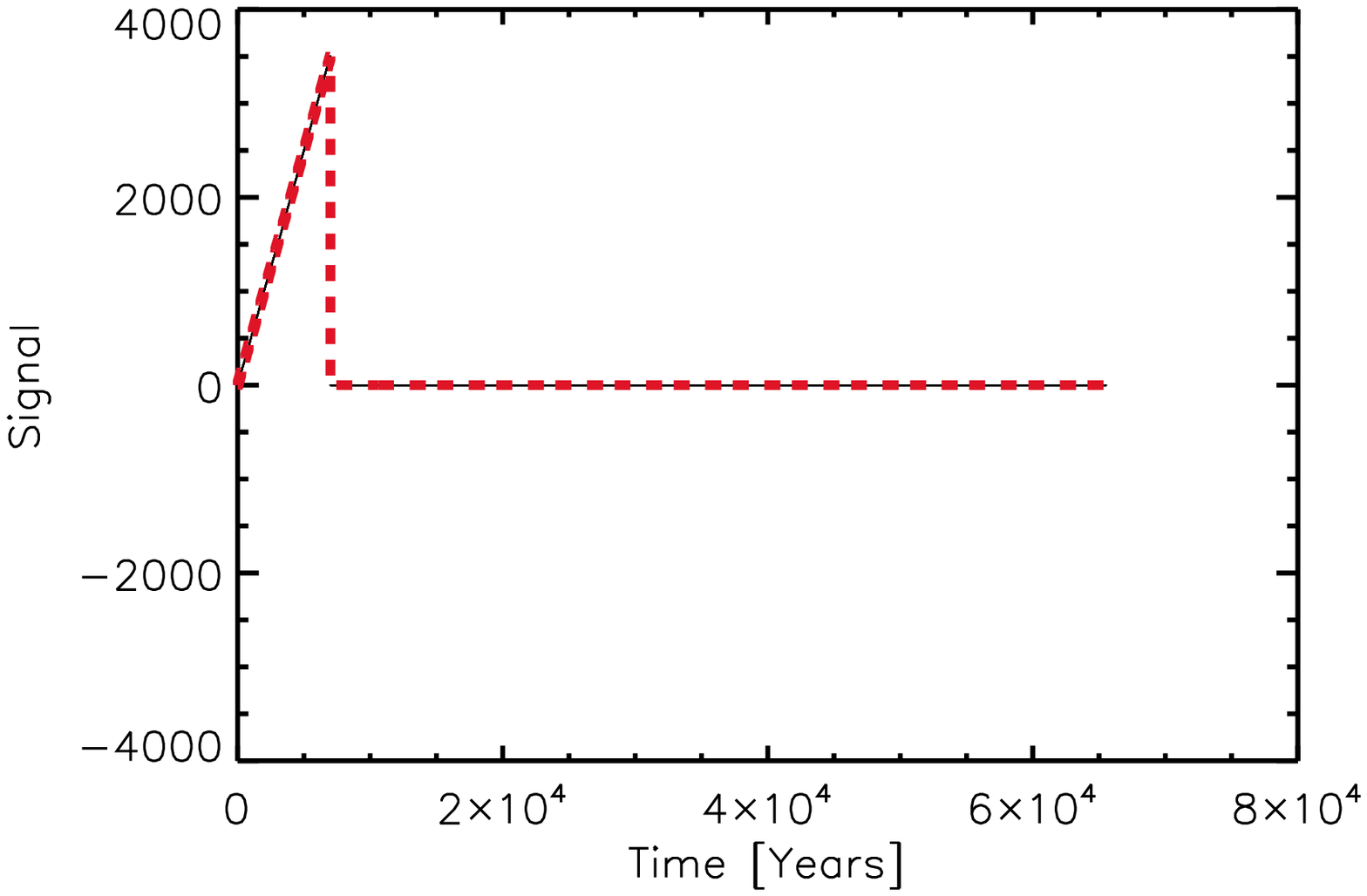}
\includegraphics[scale=0.45]{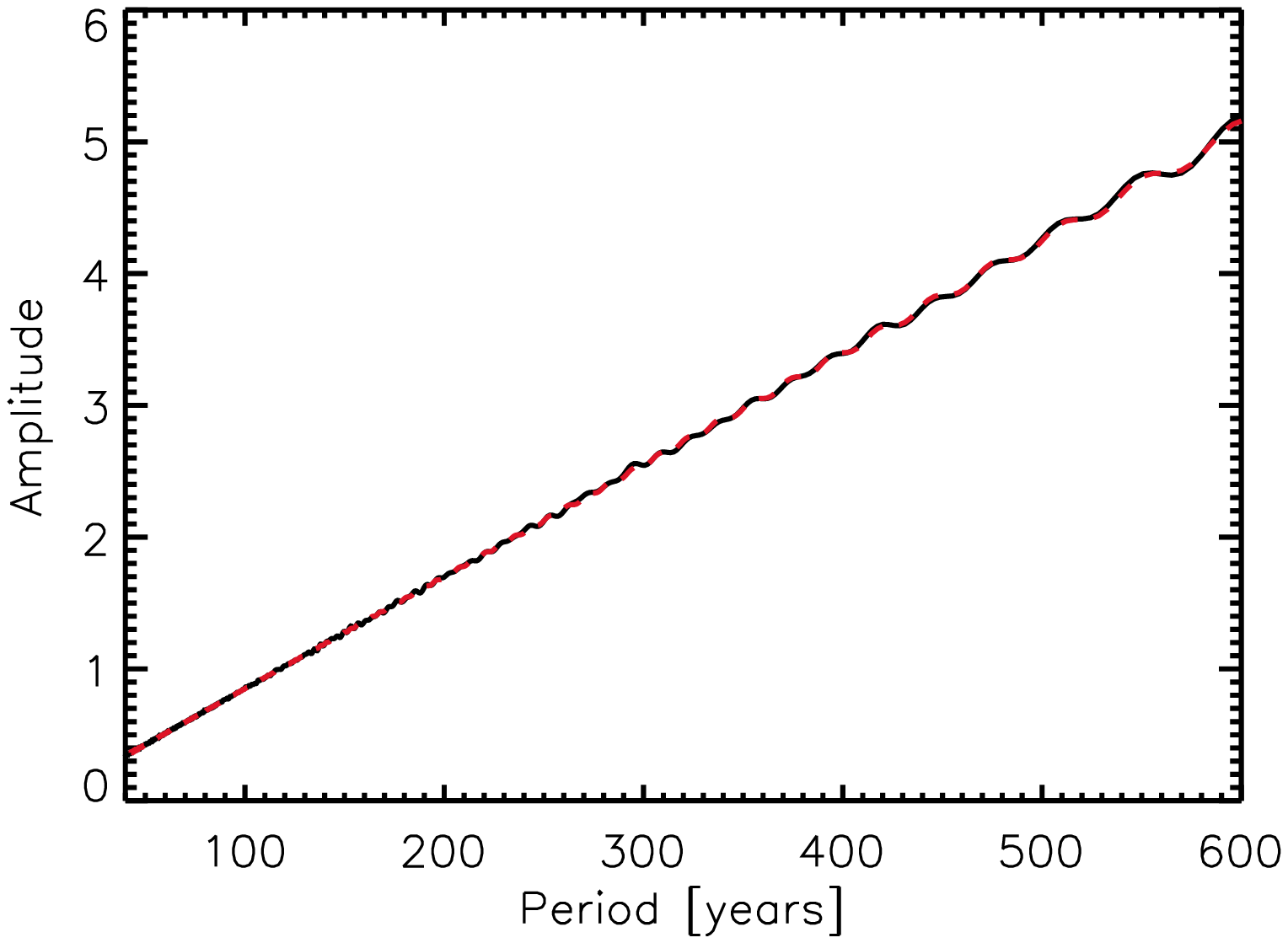}\\
\includegraphics[scale=0.45]{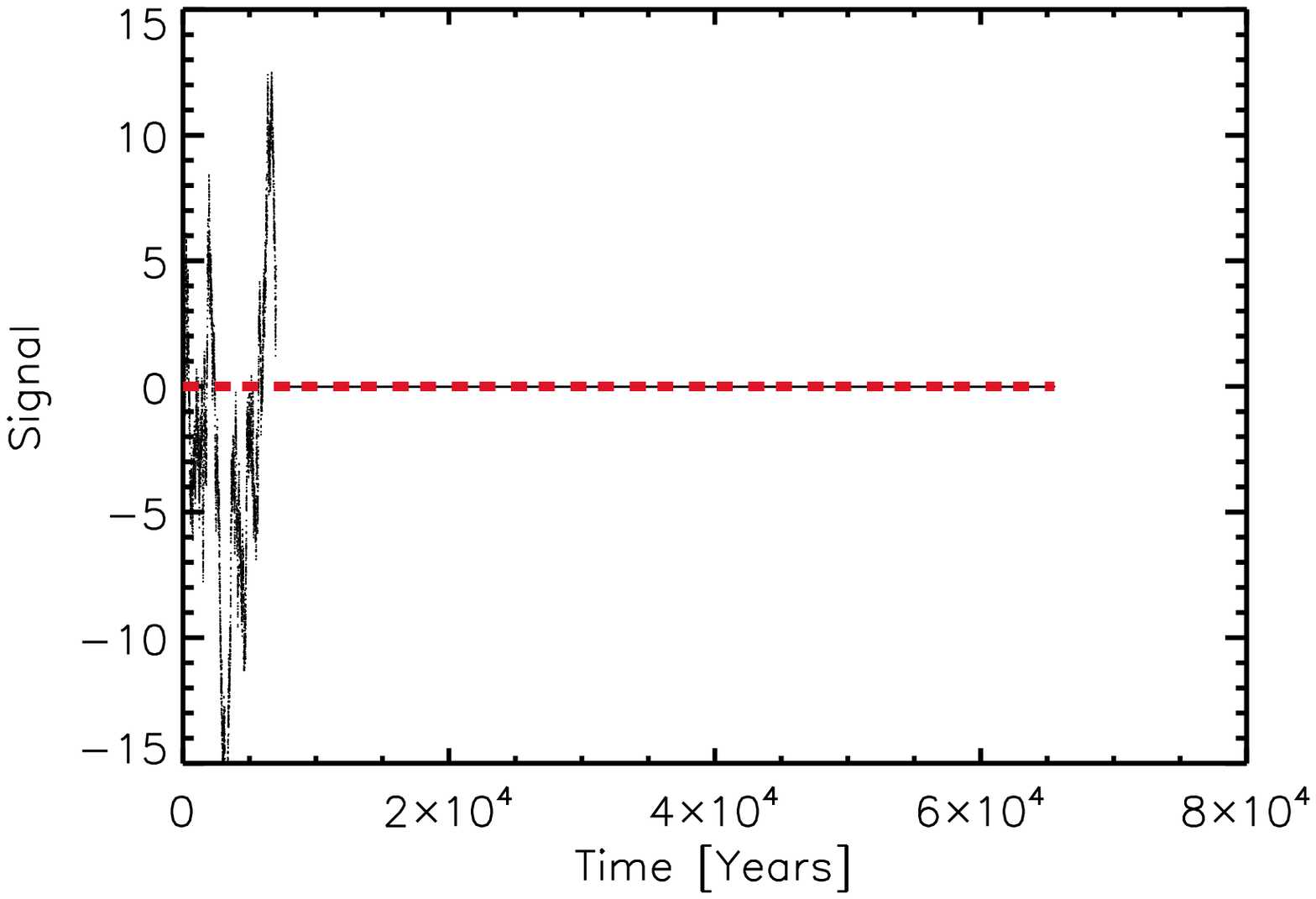}
\includegraphics[scale=0.45]{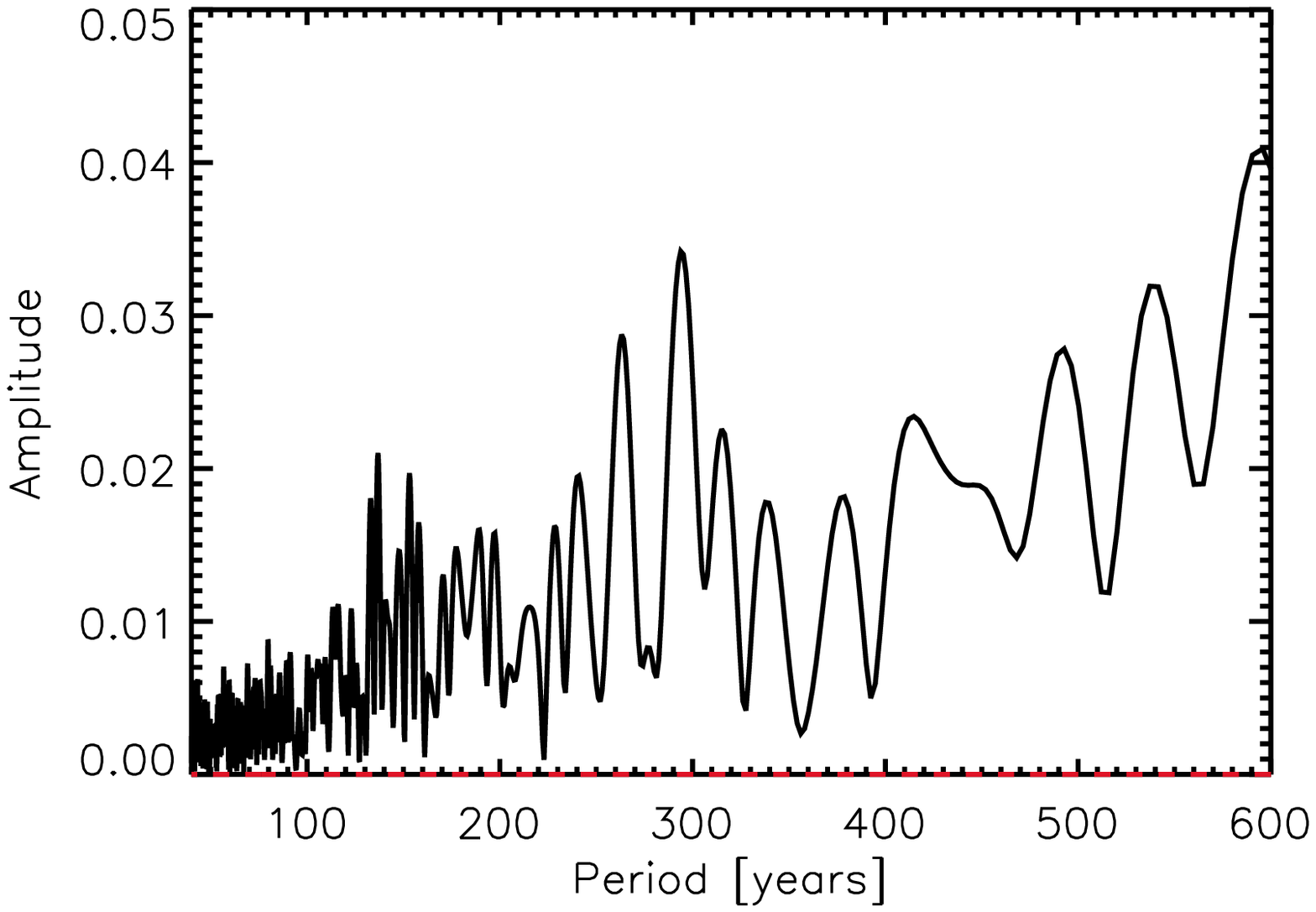}
\caption{Similar to Figure~\ref{fig:white}, but for red noise.}
\label{fig:red}
\end{center}
\end{figure*}

The random series considered by A2012 to be white noise consist of
independent, uniformly distributed, random numbers between 0 and
1. Consequently, the expectation value differs from zero, in contrast to
the definition of white noise. This has serious consequences for the
Fourier spectra since the authors zero-padded the 7000-year time series
to a length of 65,536 years and neither detrended nor apodized them
(J. Beer, private communication). As a result, a step function was
introduced into the randomly generated time series. As is well known,
this leads to spurious features in the spectrum, which invalidates the
determination of the coincidence probability.  This error carries over
to the test with red noise, which was determined by A2012 as the time
integral of their ill-defined series of white noise: since the numbers
for the latter are non-negative, the resulting `red noise' series are
monotonically increasing, followed by a big drop after zero-padding.%
\footnote{In the case that the already zero-padded white-noise signal is
integrated, the drop occurs at the end of the time series segment,
leading essentially to the same consequences.} Examples of such
time series and their Fourier transforms are shown in
Fig.~\ref{fig:white} (for `white noise') and Fig.~\ref{fig:red} (for
`red noise') along with padded time series of proper white and red
noise. The figures demonstrate that the spectra as considered by A2012
are dominated by the spurious systematic components (indicated by the
dashed lines) due to the jumps in the time series.

We quantitatively evaluated the effect of these errors on the
statistical analysis by repeating the tests in the same manner as
A2012, but with 100,000 time series each of proper white noise
(uniformly distributed between $-0.5$ and $+0.5$) and proper red noise
(determined by integration of series of proper white noise). The
resulting probabilities for one of the top-twenty spectral peaks
occuring in the windows selected by A2012 are shown in
Table~\ref{tab:noise} (first two rows of the ``Corrections'') in
comparison to the results given by A2012: the chance probabilities for
coincidences in all spectral windows increase by about a factor 200 for
white noise and by more than a factor $10^6$ for red noise, both
reaching values around $10^{-4}$.

\subsection{Error in treating the randomly generated data differently 
from the observed data}

A second error in the statistical test of A2012 is that they do not
treat the random series in the same way as they treated the proxies for
solar activity.  As mentioned above, the noise series were neither
apodized nor detrended, in contrast to the treatment of the
observational data.  Moreover, the $\phi$ data were filtered by applying
a 22-year running mean \citep{Steinhilber:etal:2012}, which was not done
in the case of the random series. Repeating the test by first applying a
22-year running average, then detrending and apodizing each realisation
of proper white and red noise, respectively, we found that the
probability of a match in all bands chosen in A2012 was increased by
additional factors $\sim$5 for white noise and $\sim$36 for red noise,
both now reaching values of the order of $10^{-3}$ (see 
Table~\ref{tab:noise}, third and fourth row of the
``Corrections''). These values are higher by about 3 order of magnitude
for white noise and by about 8 orders of magnitude for red noise than
those given by A2012.

\subsection{Error introduced by window selection}
\label{subsec:bias}

After correcting the technical errors, the probabilities of a randomly
chosen realisation matching the data, using the test of A2012, are
$5\times10^{-4}$ for white noise and $2.5\times 10^{-3}$ for red
noise. These probabilities might still be considered to be sufficiently
low to reject the null hypothesis given in Section~\ref{subsec:logic}.

However, the statistical test and, in particular, the size of the
spectral windows used in A2012 were defined a posteriori on the
basis of the data themselves. The windows were chosen so that spectral
peaks from the observations (the cosmogenic isotope series) and from the
model (the torque induced by the planets) fall within the windows. This
manifestly biases the statistical test -- a procedure that can easily
produce seemingly significant results for fully random data (see
below). An unbiased procedure would have been to first define a
criterion for a satisfactory agreement of periods (evidently a constant
multiple of the period resolution), then inspect the data for such
coincidences, and then perform the test using the above criterion for
the window size.

Table~\ref{tab:windows} shows the properties of the spectral windows
selected by A2012. Note that the windows are nonuniform in terms of the
period resolution of the Fourier spectra.  In particular, the first
window (85--89~yr) has a size of 3.7 times the period resolution,
$\Delta P=P^2/T$ ($P$: period, $T=7000\,$yr: length of dataset), of
1.08~yr for that window, thus accomodating the clearly resolved
disagreement between the periodicities detected for this period range in
the different datasets.  In contrast, the last window (503--515 yr) has
a width of only 0.32 of the period resolution in the corresponding
spectral range.

\begin{table}[h!]
\caption{Properties of the spectral windows used by A2012.}
%\begin{center}
\begin{tabular}{l c c c}
\hline
\noalign{\vspace{1mm}}
Window [yr] & $w$ [yr]$^\mathrm{a}$ & $\Delta P [yr]^\mathrm{b}$ & $w/\Delta P$
     \\[0.5ex]
\hline
\rule[3mm]{0mm}{2mm}%
\hskip 2.3mm I: 85-89\hfill&   4    &   1.08   &   3.70
       \\ [0.5ex]
%\rule[3mm]{0mm}{2mm}
\hskip 1mm II: 103-106     &   3    &   1.56   &   1.92
       \\ [0.5ex]
%\rule[3mm]{0mm}{2mm}
III: 146-151     &   5    &   3.15   &   1.59
       \\ [0.5ex]
IV: 206-210     &   4    &   6.18   &   0.65
       \\ [0.5ex]
\hskip 1mm V: 503-515     &   12   &  37.01   &   0.32
       \\ [0.5ex]
%\noalign{\vspace*{2mm}}
\hline
\end{tabular}
\begin{list}{}{}
\item[$^\mathrm{a}$] width of the spectral window
\item[$^\mathrm{b}$] period resolution (at central period of window)
\end{list}
\label{tab:windows}
%\end{center}
\end{table}

To estimate a lower bound for the effect of defining the test a
posteriori on the basis of the data, we consider windows of uniform
width (in terms of the period resolution). We choose the width as
$3.7\Delta P$, corresponding to the period bin between 85 and 89
  years defined by A2012). For this bin, the planetary forcing has a
  peak at 86 years and the solar activity has a peak at 88 years.
  Regarding the forcing peak as given, for `coincidence' we should
  allow the solar peak to fall in the interval $86 \pm 2$ years, i.e.,
  in the range between 84--88 years. A2012 choose the interval 85--89
  years, the average of the ranges assuming that either planetary
  forcing or solar activity is given.  Consequently, using windows
  narrower than 3.7 times the period resolution would lead to the
  planetary forcing peak at 86 years and the solar activity peak at 88
  years being noncoincident. Therefore, a window width of $3.7\Delta
  P$ is the minimum required for a test including all 5 period bins
  considered by A2012.

Using 100,000 realisations of white and red noise, respectively, we
calculated the probabilities for coincidences of top-twenty spectral
peaks in the windows centered on those of A2012, but with widths of
$3.7\Delta P$. The resulting probabilities for coincidences in all
windows are 7.5\% for white noise and 22\% for red noise (see
Table~\ref{tab:noise}, last two rows). These numbers show that the
test has failed, i.e., the period coincidences between isotope data
and planetary torque are statistically consisten with both white noise
and red noise.

\begin{figure}
\begin{center}
\includegraphics[width=\linewidth]{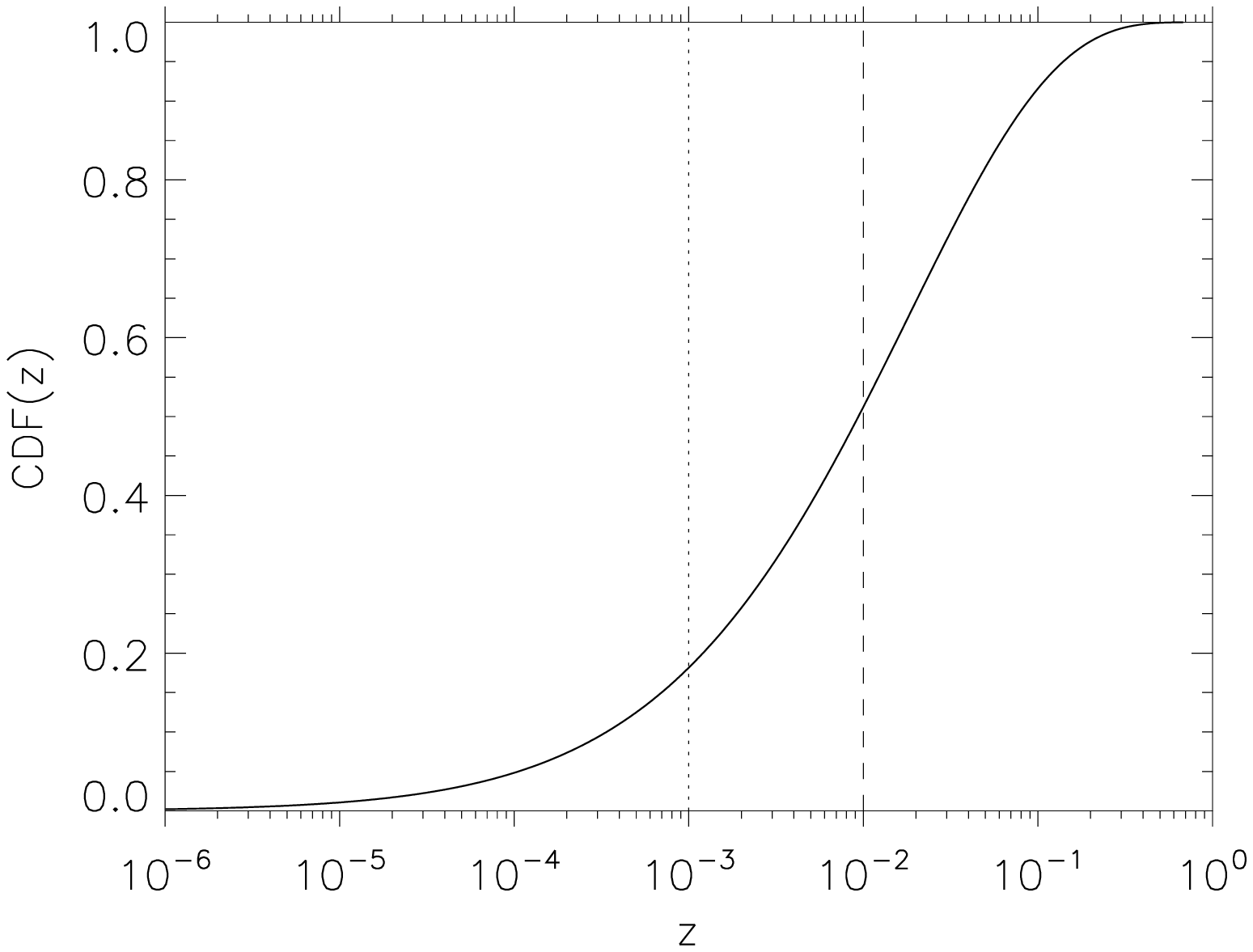}
\caption{Cumulative distribution function for the relative change
of the chance probability (for tests with random time series)
due to shrinking of the period windows. The vertical lines indicate 
apparent increases in the test significance by factors of 100 (dashed line)
and 1000 (dotted line).}
\label{fig:distr}
\end{center}
\end{figure}

The strong bias that is introduced by tayloring the widths of the
spectral windows on the basis of the data themselves can be illustrated
and quantitatively analysed as follows. Assume a set of five periods
(the `planetary' periods) and five windows in period space of width
$w_m=k\Delta P$, the value of $k$ being chosen according to a predefined
criterion of satisfactory agreement of periods. Assume a second time
series (the `solar' series) that happens to show at least one of its
top-twenty spectral peaks in each of the windows.  We can now carry out
the test with noise series and determine the probability for chance
coincidences of periods in the given windows.  It is clear that we can
make the coincidence of the periods appear more significant (i.e.,
obtain lower values for the chance coincidences in the random test) if
we inspect the data and shrink the spectral windows such that they just
cover the `planetary' peaks and the corresponding peaks of the `solar'
time series.  Assume that the `solar' time series is one realisation of
a process with a random component and that $w_m$ is sufficiently small
so that the position of the `solar' peaks within the original windows
for different realisations has a uniform distribution.%
\footnote{Actually, the same result is obtained also for a 
linear distribution in period.}
We can then easily determine the distribution of the resulting change of
the chance probability obtained in a random test. For each realisation,
the windows are shrunk according to the distance between the central
(`planetary') and the respective `solar' peak (symmetric with respect to
the central peak). The relative size change of each window is then
uniformly distributed between 0 and 1, so that the relative change
of the chance probability in a random test is given by the product of
five random numbers from this distribution. The probability density
function of this product, $z$, is given by $PDF(z)=(\ln z^4)/24$. The
corresponding cumulative distribution function ($CDF$) is shown in
Fig.~\ref{fig:distr}. It demonstrates that in about 50\% of the cases
the shrinking of the windows (biasing of the test a posteriori)
leads to an apparent increase of the significance for the random test by
a factor $>$100 and in about 18\% of the cases by a factor $>$1000. This
fits well well the results shown in Table~\ref{tab:noise}, which also
show an apparent significance boost by a factor of about 100 when going
from equal-sized windows to the hand-selected windows of A2012.

\section{`Phase locking'}

\label{subsec:locking}

\begin{figure}
\begin{center}
\includegraphics[width=\linewidth]{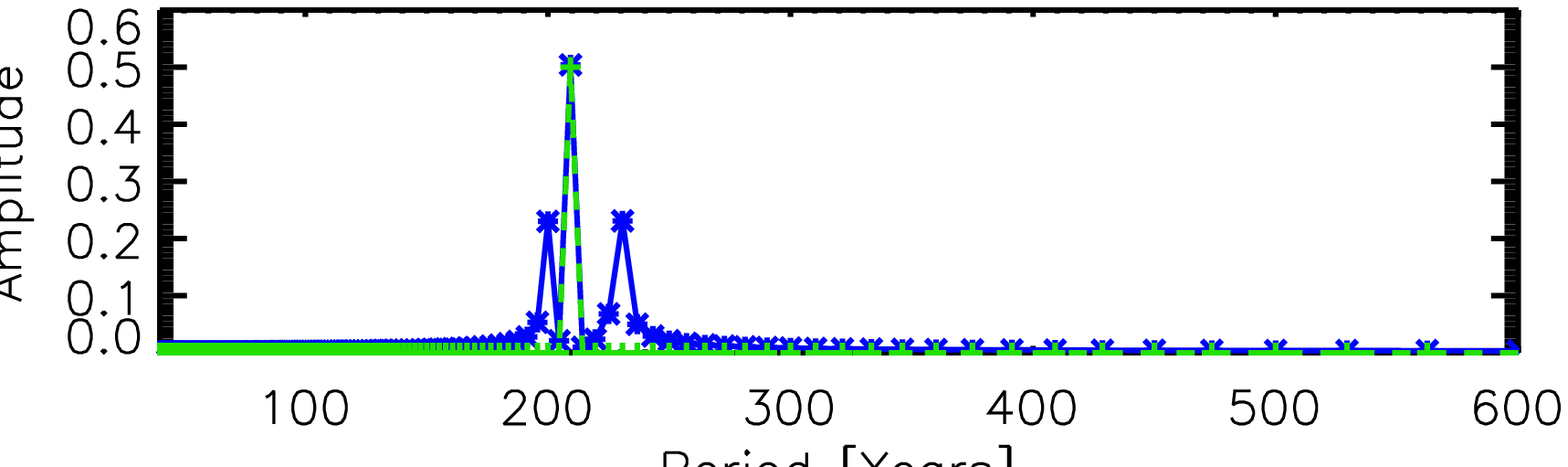}
\vskip 5mm
\includegraphics[width=\linewidth]{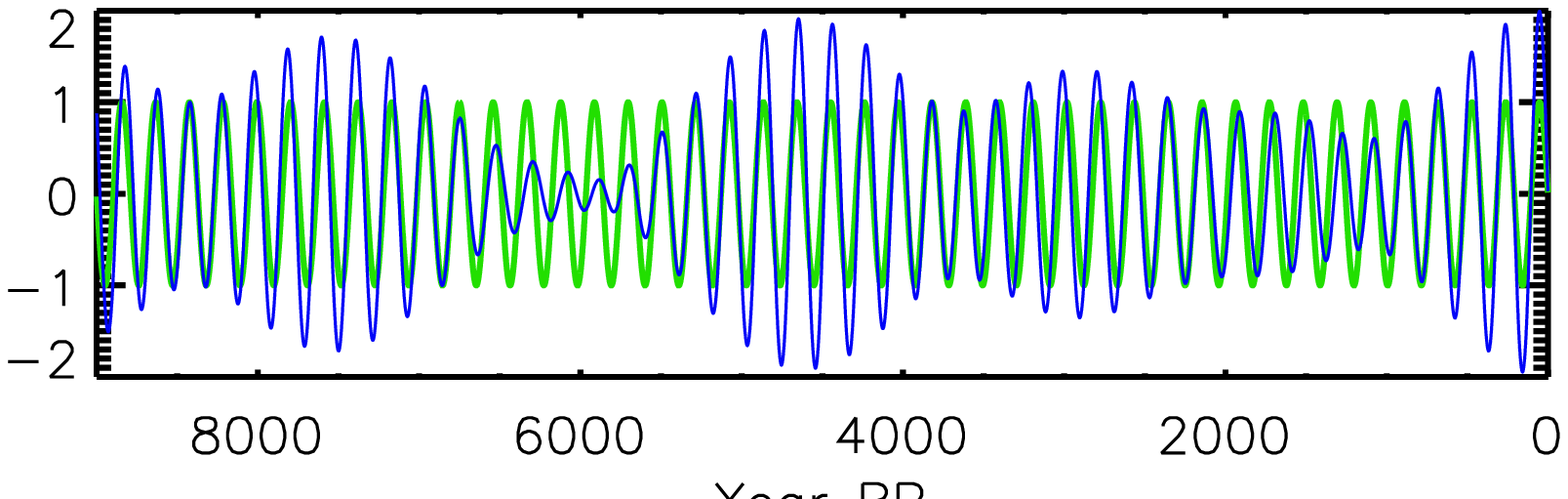}
\caption{Power spectra (upper panel) and time series (lower panel) of
  a synthetic planetary forcing signal assumed to consist of a single
  sinusoidal mode with a period of 209.3 years (green lines) and a
  synthetic solar activity signal which has two additional peaks at
  199.3 and 229.3 years. In the time domain, the side bands can be
  seen to beat against the central peak. This similar to what A2012
  interpret as phase-locking.  Since it is merely the time-domain
  counterpart to the frequency-domain signal, it does not provide any
  evidence for a physical connection of the signals.}
\label{fig:locking}
\end{center}
\end{figure}

We have seen (cf. Table~\ref{tab:windows}) that several of the period
windows selected by A2012, including that between 206 and 210 years, are
narrower than the resolution limit implied by 7000 years of data.  In
the time domain, the signal coming from these two peaks will not drift
in phase throughout the temporal interval studied.  In addition, the
bandpass filter of 190--230 years used by A2012 is broad enough to
contain side peaks with (resolvable) different frequencies and thus
introduces beating. The beating will be considerable for the solar
activity signal, which has substantial power in the side bands, but
neglible for the planetary forcing signal, which is almost monochromatic
in the frequency subdomain of the filter (cf. Fig.~5 of A2012).  Without
the effect of the beating, the two signals would not drift in
phase. Correspondingly, at times when the central peak and the side
bands are in phase, the amplitude of the (filtered) solar activity
signal is large and in phase with the planetary forcing
signal. Conversely when the components of the solar activity signal are
out of phase, the amplitude of the signal is low and the phase will be
different from that of the planetary forcing (see Figs.~6 and 7 of
A2012).

To illustrate the effect of beating on the relative phases, we
consider an example for which the planetary forcing is a simple
sinusoid with a period of 209.3 years, and the solar activity consists
of the superposition of three sinusoids with periods of 199.3, 209.3,
and 229.3 years and amplitudes of 0.5, 1., and 0.5, respectively.
Figure~\ref{fig:locking} shows this signal in the spectral and time
domains.

This demonstrates that the `phase locking' considered by A2012 as an
independent piece of evidence for an effect of the planets on solar
activity is merely a consequence of finding two peaks in a period band
narrower than the resolution limit and beating this against
sidebands. Therefore the reported `phase locking' is simply the
time-domain counterpart to the coincidence of the peaks in frequency
space and does not add to the statistical significance of the
results.

\section{Concluding remarks}

The statistical test proposed by \citet{Abreu:etal:2012}, a comparison
of the coincidences of spectral peaks from time series of planetary
torques and cosmogenic isotopes (taken as a proxy for solar activity in
the past) with red and white noise, is logically unable to substantiate
a causal relation between solar activity and planetary
orbits. Furthermore, the execution of the test contains severe technical
errors in the generation and in the treatment of the random series.
Correction of these errors and removal of the bias introduced by the
tayloring of the spectral windows a posteriori leads to probabilities
for period coincidences by chance of 22\% for red noise and 7.5\% for
white noise. The coincidences reported in \citet{Abreu:etal:2012} are
therefore consistent with both white and red noise.

Owing to our lack of understanding of the solar dynamo mechanism, red or
white noise are only one of many possible representations of its
variability in the period range between 40 and 600 years in the absence
of external effects. This is why the test of A2012 is logically
incapable of providing statistical evidence in favour of a planetary
influence.  Alternatively one could consider the probability that a
planetary system selected randomly from the set of all possible solar
systems would have periods matching those in the cosmogenic records. In
the absence of a quantitative understanding of the statistical
properties of the set of possible solar systems to draw from, the
comparison could again, at best, rule out a particular model of the
probability distribution of planetary systems. Here we have shown that
the test in A2012 does not exclude that the peaks in the range from 40
to 600 years in the planetary forcing are drawn from a distribution of
red or white noise.

We conclude that the data considered by A2012 do not provide
statistically significant evidence for an effect of the planets on solar
activity.

\acknowledgements{We are grateful to Aaron Birch for useful comments on an
earlier version of the manuscript.}

\bibliographystyle{aa.bst}
\bibliography{ms_final.bbl}

\end{document}